\documentclass[%
reprint,
showpacs,
showkeys,
nofootinbib,
nobibnotes,
amsmath,
amssymb,
aps, 
superscriptaddress,
]{revtex4-1}
\usepackage{amssymb}
\usepackage{url}
\usepackage{dcolumn}
\usepackage{bm} 
\usepackage{hyperref} 
\usepackage{siunitx}
\usepackage{textcomp} 
\usepackage{multirow}
\usepackage{tabularx} 
\usepackage{ragged2e}

\begin{document}

\title{Is DAMA Bathing in a Sea of Radioactive Argon?}

\author{D. N. McKinsey}%
\email{daniel.mckinsey@berkeley.edu}
\affiliation{
University of California Berkeley, Department of Physics, Berkeley, CA 94720, USA
}
\affiliation{
Lawrence Berkeley National Laboratory, 1 Cyclotron Rd., Berkeley, CA 94720, USA
}
\date{\today}

\begin{abstract}
A hypothesis is proposed to explain the long-standing DAMA/LIBRA puzzle. Introduced into the DAMA/LIBRA shielding is a purge gas of nominally high-purity nitrogen, which under this hypothesis contains argon impurities. Argon is introduced into the nitrogen purge gas either through leaks in the purge gas plumbing, or through commercially-supplied bottled nitrogen, diffuses through materials in the detector housings, and then comes in direct contact with the DAMA/LIBRA detectors. These argon impurities can then lead to a modulating 2.8 keV background under two scenarios. \textbf{Scenario 1):} These impurities include the isotope \textsuperscript{37}Ar, which decays by electron capture, emitting a 2.8 keV x-ray. These decays appear as single-site, monoenergetic events in DAMA/LIBRA, and produce an annual modulation due to the variation of neutron flux in the atmosphere and at the Earth's surface, which in turn leads to a seasonal variation in \textsuperscript{37}Ar production from the reactions $^{40}\textrm{Ca}(\textrm{n,}\alpha)^{37}\textrm{Ar}$ and  $^{36}\textrm{Ar}(\textrm{n,}\gamma)^{37}\textrm{Ar}$. \textbf{Scenario 2):} Radon is also in the DAMA/LIBRA purge gas, modulating seasonally at a rate below the current DAMA/LIBRA limits. When radon or its short-lived daughters decay, the resulting beta, gamma, and bremsstrahlung radiation cause stable \textsuperscript{40}Ar to be ionized within the copper housings surrounding the NaI(Tl) detectors, resulting in characteristic 2.8 keV x-rays. Modulating backgrounds might also result from radon-induced neutron or gamma-ray flux from the surrounding cavern, leading to a small modulating background enhanced at low energy by the presence of \textsuperscript{40}Ar within the copper housings. These two scenarios are straightforward to test through assay of the purge gas as well as Monte Carlo and laboratory study of the DAMA/LIBRA copper housings when excited by ionizing radiation.
\end{abstract}
\maketitle

The DAMA/LIBRA experiment is a low-background search for direct dark matter interactions using about 232 kg of high-purity NaI(Tl) crystals viewed by photomultiplier tubes, contained within a copper shield, and located deep underground in the Gran Sasso National Laboratory (LNGS). An experimental description of DAMA/LIBRA may be found in \cite{Bernabei:2008}. With the first report of a positive annual modulation signal \cite{Bernabei:1998} in 1998,  DAMA/LIBRA has ever since observed a significant seasonal variation in its event rate within an energy range of 2-6 keV electron equivalent \cite{Bernabei:2013b}. This variation is measured to be $0.0110 \pm 0.0012$ cpd/kg/keV, with a statistical significance of $9.3 \sigma$.  Many experiments have not observed concomitant dark matter interaction rates, and it is difficult to construct dark matter models that can explain DAMA yet are consistent with the various null results. Several experiments are in operation or under construction to test the DAMA/LIBRA claim.  

Numerous instrumental explanations for this annual modulation have been proffered, largely based on the variation of muon rate underground which is correlated with temperature variations in the stratosphere. These explanations include modulation of muon flux causing phosphorescence \cite{Nygren:2011}, modulation of neutrons activating \textsuperscript{128}I \cite{Ralston:2010}, or a combination of solar neutrinos and atmospheric muons \cite{Davis:2014}. These explanations all have their difficulties (see \cite{Barbeau:2014,Bernabei:2014,Klinger:2015}), largely consisting of the small muon flux underground, the single-hit nature of the DAMA/LIBRA excess, and the challenge of confining the excess to the 2-6 keV energy range.  

Here two scenarios are proposed that could explain the DAMA anomaly, both involving the generation of 2.8 keV x-rays from argon atoms in the DAMA nitrogen purge gas. The argon might enter the DAMA/LIBRA purge system through leaks in its plumbing, or simply from the tanks of commercially produced high-purity nitrogen that feed the purge system. Unfortunately, it is very common for nominally high-purity nitrogen gas to actually have substantial levels of argon contamination, as for most applications a bit of argon in the nitrogen is of no consequence. When specifying nitrogen purity, many vendors do not distinguish between noble gases and the nitrogen itself, preferring to only specify impurity levels in terms of chemically reactive impurities such as O\textsubscript{2}, H\textsubscript{2}O, CO\textsubscript{2}, CO\textsubscript{2}, and hydrocarbons.

\textbf{Scenario 1):} A model which parsimoniously explains the features of the DAMA/LIBRA excess involves the isotope \textsuperscript{37}Ar, which has a 35.04-day half-life and is produced through the reactions $^{36}\textrm{Ar}(\textrm{n},\gamma)^{37}\textrm{Ar}$ (through thermal neutron capture in the atmosphere) 
and $^{40}\textrm{Ca}(\textrm{n,}\alpha)^{37}\textrm{Ar}$ (fast neutron interactions with calcium in the soil). The latter reaction has a neutron cross-section peaking in the 5-6 MeV range, at approximately 200 mb \cite{Barnes:1974}. Upon its decay by electron capture, $^{37}\textrm{Ar} + e^{-} \rightarrow\, ^{37}\textrm{Cl} +\:\nu _{e}$, x-rays can be released from the capture of the K-shell, L-shell, and M-shell electrons at 2.8\,keV, 0.270\,keV, and 0.0175\,keV with branching ratios of 0.90, 0.09, and 0.009 \cite{Barsanov:2006}. Of interest here is the argon K-shell electron at 2.8\,keV.

Famously used \cite{Cleveland:1998} by Ray Davis and collaborators to detect the solar neutrino flux through neutrino capture on \textsuperscript{37}Cl, and later used as a neutrino calibration source for the SAGE  experiment \cite{Haxton:1988,Abdurashitov:2006},\textsuperscript{37}Ar has also been employed to calibrate gaseous neon \cite {Arnaud:2018}, two-phase Ar \cite{Sangiorgio:2013} and two-phase Xe \cite{Akimov:2014,Boulton:2017} detectors at 0.27\,keV and 2.8\,keV, as it provides a convenient, single-site, low-energy source. It has also been seen as a background in dark matter experiments. In the re-analysis of the LUX 2013 data set \cite{Akerib:2016}, a weak line was seen at 2.8 keV, consistent with \textsuperscript{37}Ar decay, and \textsuperscript{37}Ar decay events have  been seen in DarkSide data \cite{Agnes:2015}. Measurement of \textsuperscript{37}Ar may also be used for detection of underground nuclear explosions \cite{Aalseth:2011}, and its calibrated detection is well-established.

It is well known that neutron flux at the Earth's surface varies with the season, peaking in the spring or summer and falling in the winter. This variation is due to reduced atmospheric density, and in many locations, also because of winter snow cover and higher water content in the soil, which helps to moderate neutrons. The neutron flux is increased at higher altitudes, and it is known that this in turn increases the local \textsuperscript{37}Ar content \cite{Riedmann:2011} in soil air. At Pic-du-Midi in the French Pyrenees, a 25\% variation in the 0.1 to 20 MeV neutron flux is observed \cite{Cheminet:2014}, with a ratio between thermal, epithermal, and fast neutron fluxes that varies with the season. Because \textsuperscript{37}Ar is produced predominantly by neutron interactions, one may reasonably expect that its content in the atmosphere also depends on the season, with especially high production at high elevations, such as on the slopes of Gran Sasso (elevation 2912 m) and other mountains of Abruzzo. At LNGS, the well-ventilated experimental halls may be expected to have similar \textsuperscript{37}Ar levels as the surrounding countryside. If argon manages to enter the DAMA/LIBRA purge system, then \textsuperscript{37}Ar may diffuse through sealing materials in the DAMA/LIBRA detector housings and come in direct contact with its NaI(Tl) crystals, creating 2.8 keV single-site events. 

For a purge gas containing 1\% argon, an \textsuperscript{37}Ar rate of 3 mBq per liter of argon gas, 25000 cm$^2$ of NaI surface area exposed to the purge gas, and a purge gas-filled 1-cm gap between the copper housing and the NaI, one may expect a total \textsuperscript{37}Ar x-ray rate of order 40 per day entering the NaI. If this were to modulate by 25\% over the year, then this would correspond to a modulation of 10 cpd in the full 232 kg of NaI(Tl). In the 2-6 keV energy window this would be 0.01 cpd/kg/keV, comparable to the observed DAMA/LIBRA modulation. 

The key assumption behind Scenario 1 is that the purge gas can find its way into the copper housing surrounding the NaI detectors, as the 2.8 keV x-rays will not penetrate the copper assuming it is 1-2 mm thick. These copper housings were sealed at St. Gobain, encapsulating the NaI crystals so as to avoid crystal degradation due to moisture. But if the seals were to open to any degree, this might not be obvious since the DAMA/LIBRA purge gas is free of moisture. Dedicated checking of the DAMA/LIBRA detectors could test whether the copper housings have remained intact. With more technical details about the nature of the seals and any epoxy or o-rings used, one could study the permeation of argon through the seal. 

The above hypothesis may be easily tested, first by measuring the argon content of the high-purity nitrogen purge gas used in DAMA, with samples taken from the bottles themselves, from just before entering the experiment, and on the output nitrogen stream. This will help to diagnose any ways that argon might be entering the purge gas. Second, the purge gas could be measured specifically for \textsuperscript{37}Ar using facilities built specifically for this purpose \cite{Williams:2016}. When the \textsuperscript{37}Ar contamination level is established, it would be informative to perform a Monte Carlo study of the degree to which \textsuperscript{37}Ar contributes to the DAMA/LIBRA excess, using detailed and accurate models of the NaI crystals, fused silica light guides, photomultipliers, and copper housings, to calculate the fraction of \textsuperscript{37}Ar x-rays that could reach a NaI crystal. In addition, measurement of the magnitude and seasonal variation of \textsuperscript{37}Ar backgrounds in LNGS would in any case be of great interest for the overall dark matter program.

\textbf{Scenario 2):} A second possibility, also involving argon 2.8 keV x-rays, is that argon gas inside the copper housings (directly surrounding the NaI crystals) is being excited due to seasonally varying \textsuperscript{222}Rn atoms or associated neutron and gamma-ray flux, as has been seen in the Soudan laboratory \cite{Tiwari:2017}. The argon may remain from the original encapsulation of the NaI detectors if they are very well sealed, or it might come from argon impurities in the nitrogen purge gas, equilibrating through diffusion in argon content with the gas inside the copper housing. Gamma or beta rays (from e.g. radon daughters covering the copper housings) that scatter or are absorbed by passive material in the housing may also deposit some energy in the gas inside the housing, either directly or through bremsstrahlung radiation. If this gas contains some argon, then the ionized argon will emit 2.8 keV x-rays, which may then penetrate the Tetratec-teflon tape wrapping the NaI(Tl) detectors. This mechanism is similar to that used in the particle-induced x-ray emission (PIXE) method, which is widely used for measurement of trace elements \cite{Johansson:1976}. In this case the exciting particles are beta and gamma radiation emitted by \textsuperscript{222}Rn and its daughters or by neutron-activated materials, while the element being measured is \textsuperscript{40}Ar, as its characteristic 2.8 keV x-ray is readily detected by the NaI(Tl) crystals in DAMA/LIBRA.  

DAMA sets an upper limit of $5.8 \times 10^{-2}$ Bq/m\textsuperscript{3} ($\sim 5000\, \textrm{cpd/m}^3$) for \textsuperscript{222}Rn decays in its nitrogen purge gas \cite{Bernabei:2013a}, based on an analysis of double coincidences of gamma-rays from the decay of \textsuperscript{214}Bi. The \textsuperscript{222}Rn atom, with half-life 3.8 days, decays to a chain of four short-lived daughters before decaying to \textsuperscript{210}Pb which has a half-life of 22 years. If this combined rate of $\sim 25,000 \,\textrm{cpd/m}^3$ exhibits a small annual modulation, then this may be passed on to \textsuperscript{40}Ar excitation inside the copper housing, resulting in an annual modulation at 2.8 keV. The required $0.0011$ cpd/kg/keV corresponds to about 10 cpd within the 2-6 keV energy range, well within the realm of possibility given the above scenario.

This mechanism can evade the current DAMA/LIBRA background limits because even a small rate of background events, depositing most energy into passive materials and thereby evading detection at higher energies, can efficiently channel their signal into the 2-6 keV energy range through argon K-shell excitation. The limit on overall modulation across the energy spectrum above 90 keV, denoted $R_{90}$, is consistent with zero but with uncertainty ranging from 0.17 to 0.19 cpd/kg, or about 40 cpd. This limit is then weaker than the observed 10 cpd modulated event rate in the 2-6 keV energy region. So if the rate of argon x-ray production is roughly 25\% or greater than the integrated event rate above 90 keV, then these 2.8 keV x-rays could produce the observed excess in the 2-6 keV energy window. 

In principle, such decay events could also create multiple-hit events in DAMA/LIBRA; these are events in which more than one NaI detector sees an energy deposition above threshold. The rate quoted by DAMA for multiple-hit modulation is consistent with zero within the 2-6 keV window, with an uncertainty of $ \pm 0.0004$ cpd/keV/kg, or about 4\% of the 2-6 keV modulation signal. In order to create the DAMA/LIBRA excess while circumventing this limit, event types would be needed that excite argon atoms and have a probability less than 4\% of causing a simultaneous hit in a different NaI(Tl) detector. This seems plausible, given that many of these events will be from beta particles or low-energy gamma-rays, which would be unlikely to interact in more than one NaI(Tl) detector.

This second scenario doesn't require any \textsuperscript{37}Ar, just a bit of argon impurities in the gas in which the NaI crystals are sealed. The Rn decay rate, the Rn modulation amplitude, neutron modulation amplitude, amount of argon excitation, and efficiency of 2.8 keV x-ray detection may each be tuned, so that their product gives a total detection rate compatible with the DAMA result. By design, this scenario only creates argon x-ray events at 2.8 keV, since the chance of double x-ray detection (leading to events at 5.6 keV) is small by comparison. The K-shell x-rays from nitrogen, which would also be produced in this scenario, are at 0.392 keV and invisible to the DAMA/LIBRA experiment. 

The scenarios described above have other noteworthy properties. They can help to explain why, relative to a measured \textsuperscript{nat}K contamination of 13 ppb \cite{Bernabei:2012}, an increased amount (20 ppb) of \textsuperscript{nat}K is needed to fit the constant 3 keV peak that is traditionally ascribed to x-rays from \textsuperscript{40}K decay \cite{Kudryavtsev:2009}. The difference may be explained by a constant (non-modulating) argon x-ray background, which will not affect fits to data at higher energies. In addition, the observed amplitude of the modulation signal substantially decreased after the original DAMA data gathered from 1995 to 2001, for which the published modulation amplitude in the 2-6 keV bin was reported to be 0.0200 $\pm$ 0.0032 cpd/kg/keV. One can certainly expect that the amplitude of the argon x-ray modulation will be geometry-dependent, and could easily have changed between DAMA/NaI and DAMA/LIBRA. 

In conclusion, it is proposed that the annual modulation in the DAMA/LIBRA single-site event rate is due to argon gas, a Trojan horse within the copper housing surrounding the NaI(Tl) detectors. This hypothesis may be tested by monitoring of the DAMA/LIBRA purge gas for argon impurities, by measuring the purge gas and atmosphere in LNGS for \textsuperscript{37}Ar and \textsuperscript{222}Rn content and their seasonal variations, testing the copper housings for leaks and argon diffusion rates, and through studies of the x-ray production of argon-contaminated nitrogen gas when excited by ionizing radiation. Once the argon content of the purge gas has been determined, it would be informative for DAMA to use different sources of purge gas with varying levels of argon, measuring the effect of argon concentration on the event rate in the 2-6 keV energy range. Low levels of argon concentration in nitrogen may be achieved through use of boiloff gas from liquid nitrogen (see \cite{Simgen:2014} and references therein). Given a detailed model of the detector housings and all materials used in their construction, NaI(Tl) detectors within, and argon content of the enclosed gas, a Monte Carlo study of this system could be performed to simulate the decays of Rn daughters and neutron activated-materials, so as to determine whether Scenario 2 is a plausible explanation for the 2-6 keV excess.  

Useful conversations are acknowledged with Gilles Gerbier, Wick Haxton, Bill Holzapfel, Dave Nygren, Reina Maruyama, Bernard Sadoulet, and members of the LUX-ZEPLIN collaboration.

\bibliography{37Ar_bib}

\end{document}